# Symmetry-resolved two-magnon excitations in a strong spin-orbit-coupled bilayer antiferromagnet

Siwen Li[1], Elizabeth Drueke[1], Zach Porter[2], Wencan Jin[1], Zhengguang Lu[3,4], Dmitry Smirnov[3], Roberto Merlin[1], Stephen D. Wilson[2], Kai Sun[1, *], Liuyan Zhao[1, *]

[1] Department of Physics, University of Michigan, Ann Arbor, Michigan 48109

[2] Materials Department, University of California, Santa Barbara, California 93106

[3] National High Magnetic Field Laboratory, Tallahassee, Florida 32310

[4] Department of Physics, Florida State University, Tallahassee, Florida 32310

* Corresponding to: sunkai@umich.edu and lyzhao@umich.edu

We used a combination of polarized Raman spectroscopy and spin wave calculations to study magnetic excitations in the strong spin-orbit-coupled (SOC) bilayer perovskite antiferromagnet $Sr_3Ir_2O_7$. We observed two broad Raman features at $\sim 800$ cm$^{-1}$ and $\sim 1400$ cm$^{-1}$ arising from magnetic excitations. Unconventionally, the $\sim 800$ cm$^{-1}$ feature is fully symmetric ($A_{1g}$) with respect to the underlying tetragonal ($D_{4h}$) crystal lattice which, together with its broad line shape, definitively rules out the possibility of a single magnon excitation as its origin. In contrast, the $\sim 1400$ cm$^{-1}$ feature shows up in both the $A_{1g}$ and $B_{2g}$ channels. From spin wave and two-magnon scattering cross-section calculations of a tetragonal bilayer antiferromagnet, we identified the $\sim 800$ cm$^{-1}$ ($1400$ cm$^{-1}$) feature as two-magnon excitations with pairs of magnons from the zone-center $\Gamma$ point (zone-boundary van Hove singularity X point). We further found that this zone-center two-magnon scattering is unique to bilayer perovskite magnets which host an optical branch in addition to the acoustic branch, as compared to their single layer counterparts. This zone-center two-magnon mode is distinct in symmetry from the time-reversal symmetry broken "spin wave gap" and "phase mode" proposed to explain the $\sim 92$ meV ($742$ cm$^{-1}$) gap in RIXS magnetic excitation spectra of $Sr_3Ir_2O_7$.

Bilayer antiferromagnets (AFMs) of square lattice are of particular interest because they are predicted to realize a quantum phase transition from a conventional AFM phase to a long-sought quantum dimer phase [1-4] across a critical ratio ($r_c$=2.522) of nearest-neighbor interlayer ($J_c$) to intralayer ($J$) exchange coupling [5,6]. Till very recently, experimental explorations of such bilayer AFM physics have been limited to materials with very weak spin-orbit-coupling (SOC), such as AFM bilayer cuprates [7,8], manganese fluoride [9], and ruthenates [10] of perovskite structures, where $J_c$ is orders of magnitude smaller than $J$, and thus their magnetic excitations are of a simple perturbation from their single layer counterparts. The recent success in growing high-quality 5$d$ perovskite iridates with strong SOC makes it possible to have comparable $J_c$ and $J$ and result in unconventional magnetic properties.

The bilayer perovskite iridate $Sr_3Ir_2O_7$ exhibits a strong SOC-assisted Mott insulating electronic ground state [11-17] and G-type AFM order (Fig. 1a) [18-22]. Its $J_{eff}$=1/2 magnetic moment, which results from equal contributions from the three $t_{2g}$ orbitals, $d_{xy}$, $d_{xz}$, and $d_{yz}$, makes it possible for $J_c$ to be on the same order of magnitude as $J$ (Fig. 1b and 1c). The magnetic excitation spectrum of this compound, previously measured by resonant inelastic X-ray spectroscopy (RIXS), shows a prominent feature ascribed to a "spin gap" of ~92 meV (742 cm$^{-1}$) [23]. This "spin gap" is much greater than the Néel temperature energy $T_N$=285 K (~200 cm$^{-1}$) [24] and the magnetic dispersion bandwidth ~70 meV (565 cm$^{-1}$) [23], and is even comparable to the Mott charge gap of ~100 meV (807 cm$^{-1}$) [13,14,16]. As such, the magnetism in $Sr_3Ir_2O_7$ is in seemingly stark contrast to that of its single layer counterpart $Sr_2IrO_4$, a SOC cuprate analogue [25-29] whose magnetism is well described by the nearly isotropic Heisenberg spin model [30].

Two distinct theory approaches have been proposed to explain this anomalously giant "spin gap" in the RIXS data, namely, the spin wave theory [23,31] and the bond operator approach [32,33]. In the spin wave theory, the giant "spin gap" is treated as the energy cost for exciting a single zone-center magnon, which suggests an exceptionally large magnetic exchange anisotropy [23,31]. In the bond operator approach, this "gap" is ascribed to the energy of a transverse magnetic mode (*i.e.*, the phase mode), which is also a single-spin scattering process [32]. Until now, the nature of this "spin gap" in $Sr_3Ir_2O_7$ has remained elusive, except that both approaches suggest it originate from a time-reversal symmetry breaking single-spin process and require strong magnetic anisotropy. The former requires an in-depth examination as selection rules in the X-ray wavelengths is much less known than those in the optical wavelengths [34], while the latter is in direct contrast to the weak magnetic anisotropy and its associated nearly zero spin gap in the single layer counterpart $Sr_2IrO_4$.

The "spin gap" in $Sr_3Ir_2O_7$ has so far only been detected experimentally by RIXS, as the strong neutron absorption and small crystal size of iridates precludes inelastic neutron scattering as an efficient probe. Optical Raman scattering is another well-known probe for magnetic excitations in addition to phononic and electronic ones [35]. So far, Raman spectra of bilayer perovskite iridate have not revealed a signature that matches the "spin gap" [36], but show a broad continuum features centered at ~175 meV (1410 cm$^{-1}$) of $B_{2g}$ symmetry in both $Sr_2IrO_4$ and $Sr_3Ir_2O_7$ [36] arising from zone-boundary two-magnon scattering in a way similar to that in cuprates.

Here, we perform magnetic Raman measurements including symmetry channels beyond $B_{2g}$. Our temperature dependent (polarized) Raman measurements were performed in a normal incidence and backscattering geometry, the incident excitation being a CW laser with a wavelength of 532 nm (514 nm) that is focused down to ~3 $\mu$m (30 $\mu$m) in diameter at the sample site at a power of <80 $\mu$W (<1.5 mW) whereas the scattered light being analyzed by a Princeton Instrument TriVista spectrometer (a triple grating Dilor XY spectrometer).

Figure 2(a) displays temperature dependent Raman spectra taken across $T_N$ using a configuration with linearly polarized incident and unpolarized scattered light to collect as many features as possible. Three types of salient features can be immediately seen from these spectra, the sharp peaks below 700 cm$^{-1}$ present at all temperatures, the broad feature at ~800 cm$^{-1}$ only appearing at low temperatures ($M_1$, shaded in red), and the other continuum centered at ~1400 cm$^{-1}$ persisting up to room temperature ($M_2$, shaded in yellow). The sharp peaks are the Raman active optical phonons of $Sr_3Ir_2O_7$ whose frequencies are consistent with those in a previous report [35,36]. $M_2$ at low temperature exhibits a complex structure with a main broad peak at ~1400 cm$^{-1}$, two shoulders at ~1230 cm$^{-1}$ and 1300 cm$^{-1}$, and a long tail extending beyond 1700 cm$^{-1}$. This feature has been attributed to zone-boundary two-magnon scattering, whose lineshape differs from that in ref. [36] due to the different photon excitation energies. A similar feature at a similar energy was observed in the single layer counterpart $Sr_2IrO_4$, confirming that the pairs of magnons participating in this two-magnon scattering process come mainly from the in-plane Brillouin zone boundary. Finally, $M_1$, unlike $M_2$, is absent in $Sr_2IrO_4$ [36] and is the focus of this work.

We now proceed to establish the magnetic origin of $M_1$. This expectation of a magnetic origin is well-motivated by the energy scale, which matches the giant "spin gap" at ~92 meV [23,32]. Further, the temperature dependence of the $M_1$ peak intensity in Fig. 2b that closely mimics that of $M_2$, reveals an onset temperature of $T_{N'}$=230 K that coincides with the onset of the AFM order in resonant x-ray diffraction measurements [20]. In addition, the $M_1$ central frequency blueshifts by ~ 100 cm$^{-1}$ since its onset $T_{N'}$, (Fig. 2c), indicating that $M_1$ results from soft modes below $T_{N'}$. The only other possible sources of origin are phononic and electronic ones. Its greater than 100 cm$^{-1}$ linewidth precludes the possibility that it is due to a first-order single optical phonon excitation, and it is unlikely to result from any multi-phonon scattering processes because of its high intensity comparable to that of any Raman active phonons in $Sr_3Ir_2O_7$. An electronic origin can also be ruled out because, although the energy scale of ~800 cm$^{-1}$ is close to the charge gap in $Sr_3Ir_2O_7$, ~100 meV [13,14,16], this charge gap is known to be an indirect gap and should not be detected by the zero-momentum optical Raman scattering. Therefore, based on the exclusion of the phononic and electronic origins, as well as the agreement of its energy and onset temperature with those of the AFM order, we assign this broad continuum $M_1$ in $Sr_3Ir_2O_7$ arising from magnetic excitations.

Despite the fact that $M_1$ has the same energy as the giant "spin gap" from RIXS, its nature has yet to be resolved. We have performed polarized Raman spectroscopy measurements in all four selection rule channels of the underlying $D_{4h}$ crystal lattice (Fig. 3) namely, *aa*, *a'a'*, *ab*, and *a'b'* (insets of Fig. 3), corresponding to the parallel polarizations between the incident (solid arrow) and scattered (dashed arrow) polarizations aligning along and 45º rotated from *a* axis and their counterparts in the crossed channels. As expected, the optical phonons show up in only one of the $A_{1g}$, $B_{2g}$, and $B_{1g}$ symmetry channels [36], confirming the $D_{4h}$ tetragonal lattice point group. In contrast to the phonons, $M_2$ appears not only in the $B_{2g}$ channel as reported in ref. [36], but also in the $A_{1g}$ channel. This is, however, not surprising for zone-boundary two-magnon scattering, as similar observations were previously reported in cuprates [37]. Remarkably, $M_1$ can only be observed in the $A_{1g}$ channel, showing that the magnetic excitations responsible for $M_1$ preserve all symmetry operations of the underlying $D_{4h}$ lattice point group. It is known that any single-spin excitations definitely break either time reversal symmetry, corresponding to magnetism-induced circular dichroism and birefringence, or lattice point symmetries, resulting in magnetism-induced linear dichroism and birefringence [38]. Thus, the full symmetry of $M_1$, together with its broad line shape, clearly rules out the possibility that it is due to single-spin excitations assigned in a recent Raman study of $Sr_3Ir_2O_7$ [39]. In the following, we show that it originates from two-spin excitations.

We performed two-magnon scattering calculations based on the spin wave theory of a SOC bilayer Heisenberg AFM. The motivation is threefold. First, it is consistent with the fact that the broad continuum at ~ 1400 cm$^{-1}$, zone-boundary two-magnon scattering, is present in the Raman spectra of Sr$_3$Ir$_2$O$_7$ [36]. Second, it is corroborated by a recent study showing that the in-plane Ir-O-Ir length of 3.90 Å is notably smaller than the out-of-plane length of 4.06 Å (Fig. 1b and 1c) [40], suggesting that $J_c/J$ would be significantly smaller than the quantum critical point at $r_c$ = 2.522 when the orbital character of $J_{eff}$ =1/2 is nearly isotropic. Finally, the choice of this model is self-consistent in that it gives $r$=0.19 and an intralayer exchange coupling strength in Sr$_3$Ir$_2$O$_7$ comparable to that in Sr$_2$IrO$_4$. We adopt a leading order Loudon-Fleury scattering Hamiltonian, $H_{LF} = \alpha \sum_{<i,j>} (\vec{E}_I \cdot \vec{\sigma}_{ij})(\vec{E}_S \cdot \vec{\sigma}_{ij}) \vec{S}_i \cdot \vec{S}_j$, where $\vec{E}_I$ and $\vec{E}_S$ are the incident and scattered electric fields, respectively, and $\vec{\sigma}_{ij}$ is the unit vector connecting sites $i$ and $j$ [38]. We examine both a simplified spin Hamiltonian with only nearest-neighbor AFM exchange couplings and a more realistic one considering up to the third-nearest-neighbor coupling [30] and a dipole-like spin exchange for strong SOC [41]. Through their comparisons, we find that the interpretation of the fully symmetric M$_1$ as zone-center two-magnon scattering is robust in that it is independent of the choice of spin Hamiltonian.

We begin with a simple spin model with only nearest-neighbor intralayer ($J$) and interlayer ($J_c$) coupling to grasp the necessary elements for understanding M$_1$. Because the bilayer doubles the number of degrees of freedom, there are two sets of doubly degenerate magnon bands. At Γ point, one set of doubly degenerate magnon bands remains gapless with a linear dispersion (acoustic magnon branch), similar to that of Sr$_2$IrO$_4$ [30], and importantly, the other set is gapped by a finite energy (optical magnon branch), absent in Sr$_2$IrO$_4$ (Fig. 4a inset, left). Consequently, the magnon density of states (DOS) has a step-like jump at this gap energy (Fig. 4a inset, middle), and an interlayer onsite Loudon-Fleury scattering Hamiltonian leads to an observable feature of zone-center two-magnon excitations in the Raman spectra (Fig. 4a inset, right). Because Γ point is the highest symmetry point in the momentum space that preserves all of the symmetry operations of the crystal lattice, the zone-center two-magnon scattering feature should be fully symmetric with respect to the lattice [42]. Therefore, M$_1$ can be understood as two-magnon excitations with pairs of magnons from the optical branch at Γ point, with calculated energy $4\sqrt{JJ_c}$ [42]. Meanwhile, we note the difference in M$_1$ lineshape between the experimental and calculated spectra that experimental data shows a long tail whereas the calculation depicts a sharp drop at the lower-energy side of M$_1$. This difference arises mainly from two factors omitted in the calculations but present in experiment, the thermal broadening effect that is expected to impact more on the lower-energy side for a broad feature like M$_1$ and the magnon-magnon interactions that could create two-magnon states at energies lower than twice optical magnon gap.

We also confirm that M$_2$ originates from two-magnon excitations at the zone-boundary. The fact that the two magnon bands are degenerate and dispersionless along X-M (Fig. 4a inset, left) leads to a divergent DOS at every momentum point along X-M (Fig. 4a inset, middle), resulting in zone-boundary two-magnon scattering feature prominent in Raman spectra of bilayer AFMs (Fig. 4a inset, right). Due to divergent DOS, quantum corrections to account for magnon interactions are needed in computing two-magnon scattering, making the actual energy of the two-magnon feature reduced by a factor of 0.73 [43] from the directly computed value of $2\sqrt{J(4J+2J_c)}$. Note that this correction factor of 0.73 is confirmed appropriate in Sr$_2$IrO$_4$ as the ratio of its two-magnon energy in Raman (160 meV [36]) to twice its single magnon energy at X point (220 meV=2×110 meV [30]) is ~0.73. Comparing these calculated results with our experimental values of ~ 800 cm$^{-1}$ and ~1400 cm$^{-1}$, we obtain estimated values of $J$=458 cm$^{-1}$ and $J_c$=87 cm$^{-1}$. The value of $r$=$J_c$/$J$=0.19 is much smaller than the quantum critical point $r_c$=2.522, which in turn corroborates our choice of the spin wave theory in interpreting the magnetic excitations in Sr$_3$Ir$_2$O$_7$.

In reality, a more sophisticated spin Hamiltonian is needed to describe Sr$_3$Ir$_2$O$_7$ magnetism [41], including interlayer exchange coupling ($J_c$=91 cm$^{-1}$ optimized in this work), first, second, and third nearest-neighbor intralayer exchange coupling ($J$=484 cm$^{-1}$, $J_2$=−161 cm$^{-1}$ and $J_3$=121 cm$^{-1}$ directly adopted from Ref. [30] for Sr$_2$IrO$_4$) and SOC induced dipole-like exchange coupling ($H_{SOC} = \Delta \sum_{i,i',n} (\vec{S}_{i,n} \cdot \vec{\sigma}_2)(\vec{S}_{i',n} \cdot \vec{\sigma}_2)$) with Δ=16 cm$^{-1}$ based on Ref. [41] for Sr$_2$IrO$_4$) to account for the magnon dispersion along X - M (Fig. 4a, left). The optimized $J_c$ of 91 cm$^{-1}$ here is very close to that of 87 cm$^{-1}$ from the simple spin Hamiltonian above, confirming the robustness of its value, as well as the choice of spin wave theory, in Sr$_3$Ir$_2$O$_7$. Furthermore, even when these terms in addition to $J$ and $J_c$ are taken into account, the physics for M$_1$ remains exactly the same as in the simple model (Fig. 4a) because the defining feature is the presence of the gapped optical branch at the zone-center and has nothing to do with magnons at the zone-boundary. In contrast, the physics for M$_2$ requires an extension of the simple model because now the divergent DOS appears only at the van Hove singularity point X (Fig. 4a, middle) and cannot be accessed by any nearest-neighbor two-spin scattering Hamiltonian [42], which therefore, requires the next-nearest-neighbor two-spin flip processes in the Loudon-Fleury scattering formalism. Such processes, however, violate spin conservation in AFMs by flipping two spins with the same orientation to the opposite direction, and thus, are only allowed in AFMs with strong SOC that breaks the SU(2) spin rotational symmetry. In iridates, strong SOC indeed exists and manifests itself in the spin Hamiltonian through a dipole-like spin exchange term, which is known to produce spin anisotropy and determine the orientation of AFM magnetic moments [41]. Our calculations reveal that even a very small amount of SOC in the scattering Hamiltonian can lead to the observation of zone-boundary two-magnon excitations from X point in both the $A_{1g}$ and $B_{2g}$ channels (Fig. 4a, right) [42].

We further discuss the relationship between $M_1$ and the intriguing amplitude mode near the quantum critical point $r_c$=2.522 (Fig. 4b). On the one hand, the zone-center two-magnon excitations and the amplitude mode (*i.e.*, Higgs mode) share exactly the same symmetries and can both be characterized, to the leading order, by an inter-layer on-site Loudon-Fleury scattering Hamiltonian [6]. The amplitude mode is in general damped by other low-energy excitations, and therefore, results in a broad line shape similar to that of the zone-center two-magnon excitations. These similarities imply that they are simply two different manifestations of the same heavily damped object without a well-defined boundary to distinguish the two. On the other hand, the amplitude mode and the zone-center two-magnon excitation do happen at different $r$ in a bilayer AFM. The zone-center two-magnon excitations only become visible when $J_c$ is weak compared with $J$ (*i.e.*, well below $r_c$) and dissolves into the background upon increased $J_c$. In recent numeric studies [6], the amplitude mode is only underdamped and well-defined in a very small window near the quantum critical point $r_c$, and neither amplitude mode nor zone-center two-magnon mode appears for intermediate $r$ between the two regimes.

Finally, we comment on the relationship between the zone-center two-magnon excitations in the Raman spectra and the giant "spin gap" in the RIXS spectra in Ref. [23,32]. It is apparent that these two features have nearly the same energy, but sharply distinct symmetries. A trivial explanation could be that they are two different but energetically degenerate objects that happen to be captured by Raman and RIXS in a complementary way. A less trivial possibility could be that they are indeed one and the same object, which would suggest a reconsideration of the conventional selection rules. Optical Raman selection rules are well-defined based on the electric dipole approximation, which is justified in the fact that optical wavelengths are much larger than lattice constants and is further confirmed by the correct selection rules for the phonon modes. Resonant X-ray spectroscopy selection rules are less well-defined because of contributions from higher order multipolar transitions in addition to the electric dipole transitions [34]. In the electric-dipole channel, the polarization of the incoming X-ray rotating to a perpendicular polarization in the scattered X-ray reflects time-reversal-symmetry-breaking excitations, whereas in the electric-quadrupole or higher multipole channels, such a rotation is naturally allowed even for excitations of the $A_{1g}$ symmetry, as the case for the RIXS on the "spin gap" of $Sr_3Ir_2O_7$.

**Acknowledgments**

L. Zhao acknowledges support by National Science Foundation CAREER Award No. DMR-1749774. K. Sun acknowledges support by National Science Foundation Award No. NSF-EFMA-1741618. S. D. Wilson acknowledges support by National Science Foundation Award No. DMR-1905801 and by Army Research Office Award W19NF-16-1-0361 (Z. P.). E. Drueke acknowledges support by the National Science Foundation Graduate Research Fellowship Program under Grant No. DGE-1256260. The work done at National High Magnetic Field Laboratory is supported by National Science Foundation Award No. DMR-1644779.

**References**

[1] A. V. Chubukov and D. K. Morr, Physical Review B **52**, 3521 (1995).

[2] A. W. Sandvik, A. V. Chubukov, and S. Sachdev, Physical Review B **51**, 16483 (1995).

[3] C. N. A. van Duin and J. Zaanen, Physical Review Letters **78**, 3019 (1997).

[4] V. N. Kotov, O. Sushkov, Z. Weihong, and J. Oitmaa, Physical Review Letters **80**, 5790 (1998).

[5] L. Wang, K. S. D. Beach, and A. W. Sandvik, Physical Review B **73**, 014431 (2006).

[6] M. Lohöfer, T. Coletta, D. G. Joshi, F. F. Assaad, M. Vojta, S. Wessel, and F. Mila, Physical Review B **92**, 245137 (2015).

[7] A. Millis and H. Monien, Physical Review B **50**, 16606 (1994).

[8] D. Reznik, P. Bourges, H. Fong, L. Regnault, J. Bossy, C. Vettier, D. Milius, I. Aksay, and B. Keimer, Physical Review B **53**, R14741 (1996).

[9] C. van Uijen, A. Arts, and H. de Wijn, Solid state communications **47**, 455 (1983).

[10] X. Ke, T. Hong, J. Peng, S. Nagler, G. Granroth, M. Lumsden, and Z. Mao, Physical Review B **84**, 014422 (2011).

[11] S. J. Moon *et al.*, Physical Review Letters **101**, 226402 (2008).

[12]     B. M. Wojek, M. H. Berntsen, S. Boseggia, A. T. Boothroyd, D. Prabhakaran, D. F. McMorrow, H. M. Rønnow, J. Chang, and O. Tjernberg, Journal of Physics: Condensed Matter **24**, 415602 (2012).

[13]     Y. Okada *et al.*, Nature Materials **12**, 707 (2013).

[14]     Q. Wang, Y. Cao, J. A. Waugh, S. R. Park, T. F. Qi, O. B. Korneta, G. Cao, and D. S. Dessau, Physical Review B **87**, 245109 (2013).

[15]     A. de la Torre *et al.*, Physical Review Letters **113**, 256402 (2014).

[16]     H. J. Park, C. H. Sohn, D. W. Jeong, G. Cao, K. W. Kim, S. J. Moon, H. Jin, D.-Y. Cho, and T. W. Noh, Physical Review B **89**, 155115 (2014).

[17]     H. Chu, L. Zhao, A. de la Torre, T. Hogan, S. D. Wilson, and D. Hsieh, Nature Materials **16**, 200 (2017).

[18]     I. Nagai, Y. Yoshida, S. I. Ikeda, H. Matsuhata, H. Kito, and M. Kosaka, Journal of Physics: Condensed Matter **19**, 136214 (2007).

[19]     S. Boseggia, R. Springell, H. C. Walker, A. T. Boothroyd, D. Prabhakaran, S. P. Collins, and D. F. McMorrow, Journal of Physics: Condensed Matter **24**, 312202 (2012).

[20]     S. Boseggia, R. Springell, H. C. Walker, A. T. Boothroyd, D. Prabhakaran, D. Wermeille, L. Bouchenoire, S. P. Collins, and D. F. McMorrow, Physical Review B **85**, 184432 (2012).

[21]     C. Dhital *et al.*, Physical Review B **86**, 100401 (2012).

[22]     S. Fujiyama, K. Ohashi, H. Ohsumi, K. Sugimoto, T. Takayama, T. Komesu, M. Takata, T. Arima, and H. Takagi, Physical Review B **86**, 174414 (2012).

[23]     J. Kim *et al.*, Physical Review Letters **109**, 157402 (2012).

[24]     J. W. Kim, Y. Choi, J. Kim, J. F. Mitchell, G. Jackeli, M. Daghofer, J. van den Brink, G. Khaliullin, and B. J. Kim, Physical Review Letters **109**, 037204 (2012).

[25]     B. J. Kim *et al.*, Physical Review Letters **101**, 076402 (2008).

[26]     Y. K. Kim, O. Krupin, J. D. Denlinger, A. Bostwick, E. Rotenberg, Q. Zhao, J. F. Mitchell, J. W. Allen, and B. J. Kim, Science **345**, 187 (2014).

[27]     Y. K. Kim, N. H. Sung, J. D. Denlinger, and B. J. Kim, Nature Physics **12**, 37 (2015).

[28]     L. Zhao, D. H. Torchinsky, H. Chu, V. Ivanov, R. Lifshitz, R. Flint, T. Qi, G. Cao, and D. Hsieh, Nature Physics **12**, 32 (2015).

[29]     Y. Cao *et al.*, Nature Communications **7**, 11367 (2016).

[30]     J. Kim *et al.*, Physical Review Letters **108**, 177003 (2012).

[31]     X. Lu, D. E. McNally, M. Moretti Sala, J. Terzic, M. H. Upton, D. Casa, G. Ingold, G. Cao, and T. Schmitt, Physical Review Letters **118**, 027202 (2017).

[32]	M. Moretti Sala *et al.*, Physical Review B **92**, 024405 (2015).

[33]	T. Hogan, R. Dally, M. Upton, J. P. Clancy, K. Finkelstein, Y.-J. Kim, M. J. Graf, and S. D. Wilson, Physical Review B **94**, 100401 (2016).

[34]	L. J. P. Ament, M. van Veenendaal, T. P. Devereaux, J. P. Hill, and J. van den Brink, Reviews of Modern Physics **83**, 705 (2011).

[35]	W. Jin *et al.*, Physical Review B **99**, 041109 (2019).

[36]	H. Gretarsson, N. H. Sung, M. Höppner, B. J. Kim, B. Keimer, and M. Le Tacon, Physical Review Letters **116**, 136401 (2016).

[37]	R. R. P. Singh, P. A. Fleury, K. B. Lyons, and P. E. Sulewski, Physical Review Letters **62**, 2736 (1989).

[38]	P. A. Fleury and R. Loudon, Physical Review **166**, 514 (1968).

[39]	J. Zhang *et al.*, npj Quantum Materials **4**, 23 (2019).

[40]	T. Hogan, L. Bjaalie, L. Zhao, C. Belvin, X. Wang, C. G. Van de Walle, D. Hsieh, and S. D. Wilson, Physical Review B **93**, 134110 (2016).

[41]	G. Jackeli and G. Khaliullin, Physical Review Letters **102**, 017205 (2009).

[42]	See Supporting Information at xx for spin-wave dispersion and two-magnon cross section calculations for a bilayer AFM.

[43]	C. M. Canali and S. M. Girvin, Physical Review B **45**, 7127 (1992).

**Figures and Figure Captions**

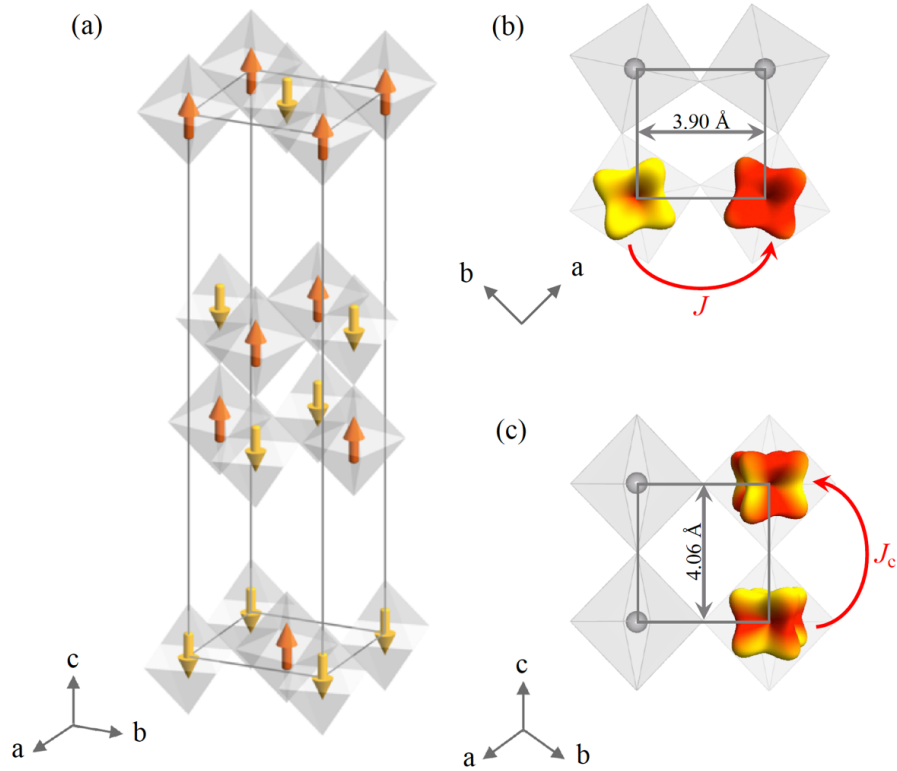

**Figure 1.** Crystalline and magnetic structures of $Sr_3Ir_2O_7$. (a) The expanded unit cell for the bilayer AFM $Sr_3Ir_2O_7$, where gray octahedra are the oxygen octahedra, gray spheres are the Ir atoms, and orange/yellow arrows are $J_{eff}=\pm1/2$ magnetic moments. (b) Top view of a layer of IrO cages, where $J$ stands for the nearest-neighbor intralayer exchange coupling. (c) Side view of two layers of IrO cages within a bilayer, where $J_c$ stands for the nearest-neighbor interlayer exchange coupling. a, b, and c are crystal axes. The yellow-red colored patterns in (b) and (c) are for the $J_{eff}=\pm1/2$ wavefunctions.

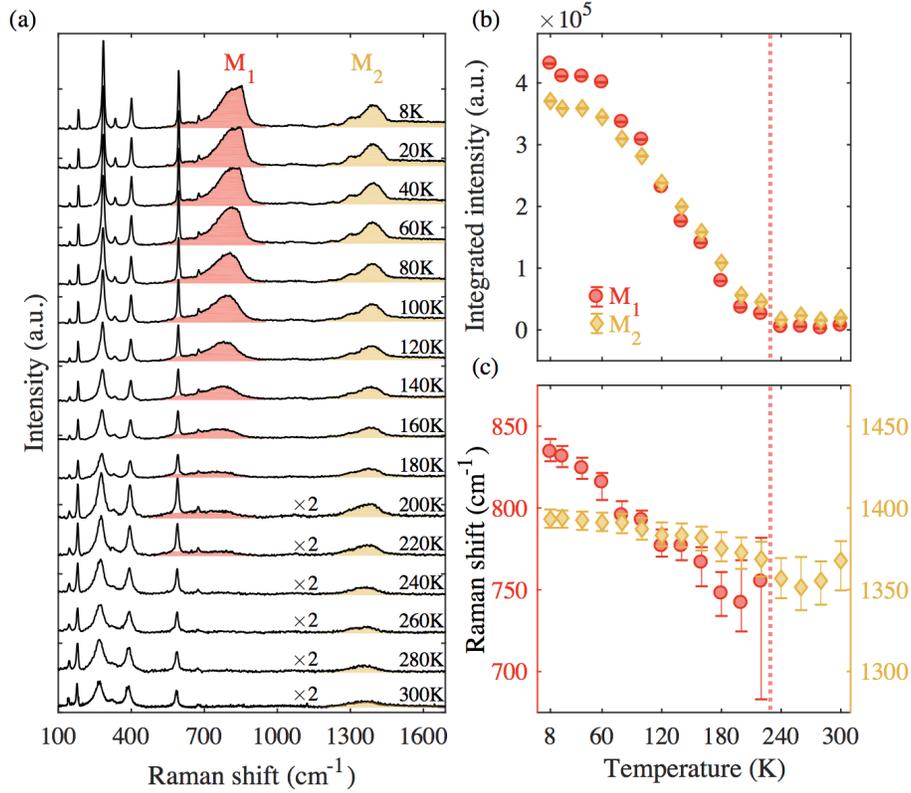

**Figure 2.** Temperature dependent magnetic Raman spectra of $Sr_3Ir_2O_7$. (a) Raman spectra taken over a temperature range from 300 K down to 8 K with linearly polarized incident (whose polarization is at about 45 degrees from crystal axis a) and unpolarized scattered light, where the spectra above 200 K are multiplied by a factor of 2. These spectra are offset vertically for clarity. $M_1$ and $M_2$ label the two broad continuums shaded in red and yellow, respectively. (b) Temperature dependence of the extracted peak intensities for $M_1$ and $M_2$. (c) Temperature dependence of the extracted central frequencies for $M_1$ and $M_2$. The error bars in (b) and (c) are defined by one standard error for the extracted parameters.

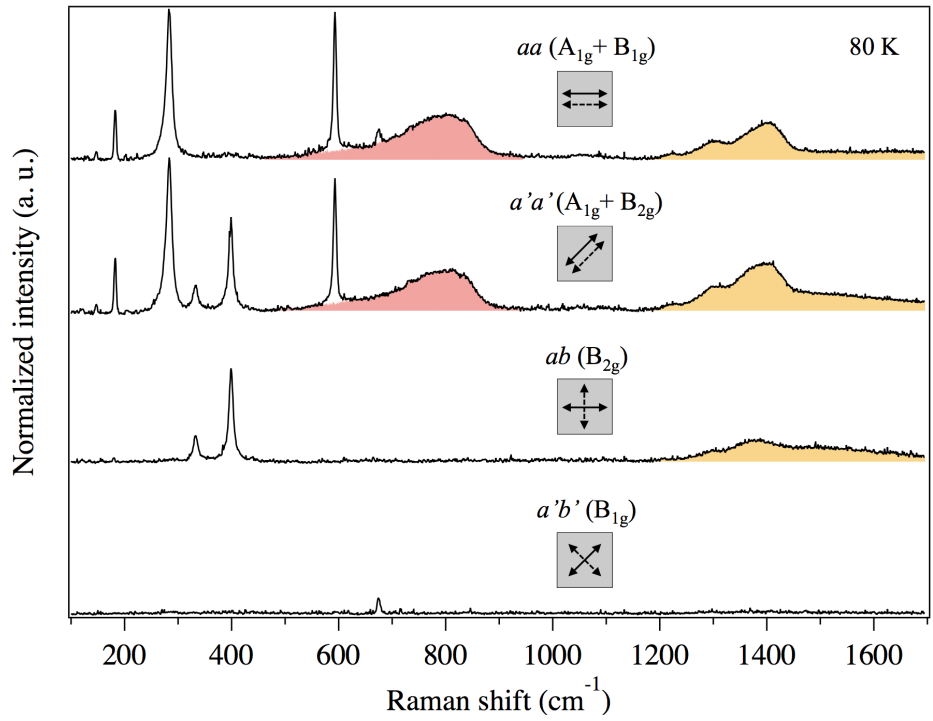

**Figure 3.** Symmetry selection rules for the magnetic excitations in $Sr_3Ir_2O_7$. Raman spectra were taken at 80 K in four polarization channels: *aa*, *a'a'*, *ab*, and *a'b'*. The insets indicate the polarization channels and the selected symmetry modes under the $D_{4h}$ point group

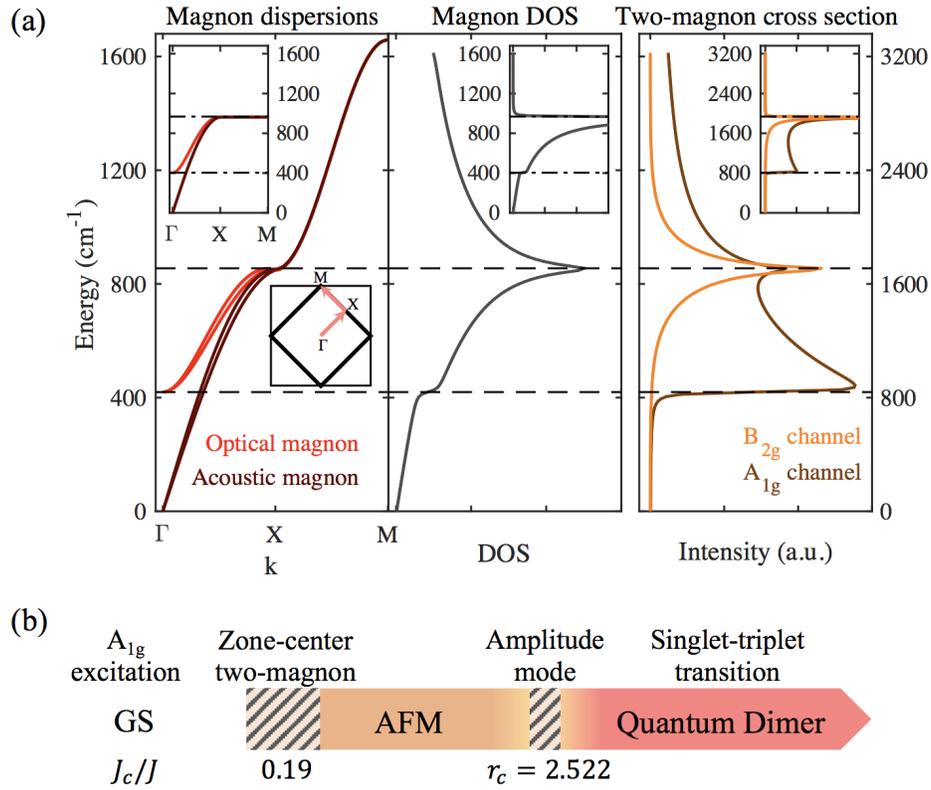

**Figure 4.** Theoretical calculations for magnetic excitations in $Sr_3Ir_2O_7$. (a) Calculated magnon band dispersions along the high symmetry cut in the momentum space $\Gamma-X-M$ (left), magnon DOS (middle), and two-magnon scattering cross section in the $A_{1g}$ and $B_{2g}$ channels (right) under the nearest-neighbor (inset) and the longer-range (main panels) exchange coupling approximation. The magnon dispersions and DOS share the same vertical axis (left side axis) while the two-magnon cross section has its own vertical axis (right side axis). (b) Schematics to illustrate the magnetic ground states and two-spin excitations as a function of $r=J_c/J$.

# Symmetry-resolved two-magnon excitations in a strong spin-orbit-coupled bilayer antiferromagnet

Siwen Li[1], Elizabeth Drueke[1], Zach Porter[2], Wencan Jin[1], Zhengguang Lu[3,4], Dmitry Smirnov[3], Roberto Merlin[1], Stephen D. Wilson[2], Kai Sun[1, *], Liuyan Zhao[1, *]

[1] Department of Physics, University of Michigan, Ann Arbor, Michigan 48109, USA

[2] Materials Department, University of California, Santa Barbara, California 93106, USA

[3] National High Magnetic Field Laboratory, Tallahassee, Florida 32310, USA

[4] Department of Physics, Florida State University, Tallahassee, Florida 32310, USA

* Corresponding authors:
Kai Sun
**Email:** sunkai@umich.edu

Liuyan Zhao
**Email:** lyzhao@umich.edu

**Table of Contents**

## 1. Calculations of the nearest-neighbor Heisenberg model of a bilayer antiferromagnet

### 1.a Magnon dispersion

The spin Hamiltonian of a bilayer magnetic system is

$$H = J\sum_{<i,j>,n}\vec{S}_{i,n} \cdot \vec{S}_{j,n} + J_c\sum_i \vec{S}_{i,1} \cdot \vec{S}_{i,2}, \qquad (1)$$

where $J$ and $J_c$ are nearest-neighbor (NN) intra- and inter-layer exchange interactions, respectively, and $\vec{S}_{i,n}$ denotes the spin operator on site $i$ in layer $n$ ($n = 1, 2$). $J > 0$ for intra-layer antiferromagnets (AFM), and $J_c > 0$ for interlayer AFMs.

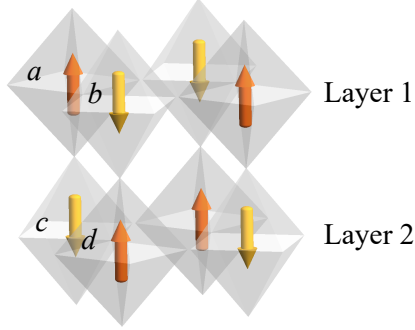

**Fig. S1** Four different sites $a$, $b$, $c$, and $d$ within one magnetic unit cell of $Sr_3Ir_2O_7$. The $a$ and $d$ sites are spin up, while the $b$ and $c$ sites are spin down.

As shown in Fig. S1, there are four different spin sites in $Sr_3Ir_2O_7$ marked as $a$, $b$, $c$ and $d$. Through the Dyson-Maleev transformation (up to the leading order) followed by a Fourier transform, we get

$$\begin{aligned}
S_{i,1}^+ &= S_{i,1}^x + iS_{i,1}^y = \sqrt{2S/N}\sum_{\vec{k}} a_{\vec{k}} e^{-i\vec{k}\cdot\vec{r}_i}, S_{i,1}^- = S_{i,1}^x - iS_{i,1}^y = \sqrt{2S/N}\sum_{\vec{k}} a_{\vec{k}}^+ e^{i\vec{k}\cdot\vec{r}_i} \\
S_{j,1}^+ &= S_{j,1}^x + iS_{j,1}^y = \sqrt{2S/N}\sum_{\vec{k}} b_{\vec{k}}^+ e^{i\vec{k}\cdot\vec{r}_j}, S_{j,1}^- = S_{j,1}^x - iS_{j,1}^y = \sqrt{2S/N}\sum_{\vec{k}} b_{\vec{k}} e^{-i\vec{k}\cdot\vec{r}_j} \\
S_{i,2}^+ &= S_{i,2}^x + iS_{i,2}^y = \sqrt{2S/N}\sum_{\vec{k}} c_{\vec{k}}^+ e^{i\vec{k}\cdot\vec{r}_i}, S_{i,2}^- = S_{i,2}^x - iS_{i,2}^y = \sqrt{2S/N}\sum_{\vec{k}} c_{\vec{k}} e^{-i\vec{k}\cdot\vec{r}_i} \\
S_{j,2}^+ &= S_{j,2}^x + iS_{j,2}^y = \sqrt{2S/N}\sum_{\vec{k}} d_{\vec{k}} e^{-i\vec{k}\cdot\vec{r}_j}, S_{j,2}^- = S_{j,2}^x - iS_{j,2}^y = \sqrt{2S/N}\sum_{\vec{k}} d_{\vec{k}}^+ e^{i\vec{k}\cdot\vec{r}_j}
\end{aligned} \qquad (2)$$

where $S^+$ and $S^-$ are the spin ladder operators, $S^x$ and $S^y$ are the $x$- and $y$-components of the spins, $S = 1/2$ is the total effective angular momentum of Ir atoms, $N$ is the total number of primitive cells, $a_{\vec{k}}$, $b_{\vec{k}}$, $c_{\vec{k}}$, and $d_{\vec{k}}$ are bosonic operators with momentum $\vec{k}$, and $\vec{r}_i$ is the position vector of lattice site $i$. Following the transformations in Eq. (2), the spin Hamiltonian in Eq. (1) is rewritten using bosonic creation and annihilation operators in the momentum space as

$$\begin{aligned}
H = JSz\sum_{\vec{k}}[\gamma_{\vec{k}}(a_{\vec{k}}b_{\vec{k}} + c_{\vec{k}}d_{\vec{k}} + a_{\vec{k}}^+ b_{\vec{k}}^+ + c_{\vec{k}}^+ d_{\vec{k}}^+) + (a_{\vec{k}}^+ a_{\vec{k}} + b_{\vec{k}}^+ b_{\vec{k}} + c_{\vec{k}}^+ c_{\vec{k}} + d_{\vec{k}}^+ d_{\vec{k}})] \\
+ J_c S\sum_{\vec{k}}(a_{\vec{k}}c_{\vec{k}} + b_{\vec{k}}d_{\vec{k}} + a_{\vec{k}}^+ c_{\vec{k}}^+ + b_{\vec{k}}^+ d_{\vec{k}}^+ + a_{\vec{k}}^+ a_{\vec{k}} + b_{\vec{k}}^+ b_{\vec{k}} + c_{\vec{k}}^+ c_{\vec{k}} + d_{\vec{k}}^+ d_{\vec{k}})
\end{aligned} \qquad (3)$$

where $z = 4$ is the coordination number and $\gamma_{\vec{k}} = (\cos k_x + \cos k_y)/2$. This Hamiltonian can then be diagonalized via the Bogoliubov transformation,

$$\begin{pmatrix} \alpha \\ \beta \\ \gamma^+ \\ \delta^+ \end{pmatrix} = \begin{pmatrix} \frac{u_1+v_1}{\sqrt{8u_1v_1}} & \frac{u_1+v_1}{\sqrt{8u_1v_1}} & \frac{-u_1+v_1}{\sqrt{8u_1v_1}} & \frac{-u_1+v_1}{\sqrt{8u_1v_1}} \\ \frac{u_2+v_2}{\sqrt{8u_2v_2}} & \frac{-u_2-v_2}{\sqrt{8u_2v_2}} & \frac{u_2-v_2}{\sqrt{8u_2v_2}} & \frac{-u_2+v_2}{\sqrt{8u_2v_2}} \\ \frac{-u_1+v_1}{\sqrt{8u_1v_1}} & \frac{-u_1+v_1}{\sqrt{8u_1v_1}} & \frac{u_1+v_1}{\sqrt{8u_1v_1}} & \frac{u_1+v_1}{\sqrt{8u_1v_1}} \\ \frac{u_2-v_2}{\sqrt{8u_2v_2}} & \frac{-u_2+v_2}{\sqrt{8u_2v_2}} & \frac{u_2+v_2}{\sqrt{8u_2v_2}} & \frac{-u_2-v_2}{\sqrt{8u_2v_2}} \end{pmatrix} \begin{pmatrix} a \\ d \\ b^+ \\ c^+ \end{pmatrix} \quad (4)$$

where $u_1 = \sqrt{Jz(1-\gamma_{\vec{k}})}$, $v_1 = \sqrt{2J_c + Jz(1+\gamma_{\vec{k}})}$, $u_2 = \sqrt{Jz(1+\gamma_{\vec{k}})}$, and $v_2 = \sqrt{2J_c + Jz(1-\gamma_{\vec{k}})}$, which transforms these bosonic operators into magnon creation or annihilation operators.

After the diagonalization with the $4 \times 4$ matrix in Eq. (4), the eigen-energies of the magnons are calculated to be

$$E_1 = Su_1v_1 \text{ and } E_2 = Su_2v_2 \quad (5)$$

Each set of energy bands is doubly degenerate with one spin-up and one spin-down magnon branch. As shown in the insets of Fig. 4 in the main text, the optical branch gapped at the $\Gamma$ point is unique to the bilayer and contributes to a jump in the density of states (DOS) spectra at ~ 400 cm$^{-1}$. This then leads to the observation of a zone-center two-magnon feature M$_1$ at ~ 800 cm$^{-1}$. $E_1$ and $E_2$ are degenerate along $X - M$ and both contribute to the diverging DOS at ~ 950 cm$^{-1}$, which is the origin of the zone-boundary two-magnon feature M$_2$ at ~ 1900 cm$^{-1}$. (Please note that the quantum correction factor needs to be considered before comparing to the experimental value of M$_2$).

### 1.b Two-magnon cross section

The interactions between light and two-magnon excitations can be described by the Hamiltonian

$$H^S = \alpha \sum_{<i,j>,n} \phi_{ij} \vec{S}_{i,n} \cdot \vec{S}_{j,n} \quad (6)$$

which includes all symmetry-allowed combinations of the NN spin products.

For the A$_{1g}$ symmetry channel, $\phi_{ij} = 1$ for all NN sites, which preserves all symmetry operations in the D$_{4h}$ point group (Fig. S2(a)). After transforming to the magnon operators, quadratic terms of two creation or annihilation operators make dominant contributions to the two-magnon scattering we investigate here [1], which is written as

$$H^S_{A_{1g}} = \Gamma_1 \sum_{\vec{k}} (\alpha_{\vec{k}} \gamma_{\vec{k}} + \alpha^+_{\vec{k}} \gamma^+_{\vec{k}}) + \Gamma_2 \sum_{\vec{k}} (\beta_{\vec{k}} \delta_{\vec{k}} + \beta^+_{\vec{k}} \delta^+_{\vec{k}}) \quad (7)$$

where $\Gamma_1 = -J_c Su_1/v_1$ and $\Gamma_2 = J_c Su_2/v_2$.

For the $B_{2g}$ symmetry channel, $\phi_{ij} = \pm 1$ depending on the direction of the NN bond, as illustrated in Fig. S2(b). This $B_{2g}$ scattering Hamiltonian written in terms of magnon operators takes the form

$$H^S_{B_{2g}} = \Gamma_3 \sum_{\vec{k}} (\alpha_{\vec{k}} \gamma_{\vec{k}} + \alpha^+_{\vec{k}} \gamma^+_{\vec{k}}) + \Gamma_4 \sum_{\vec{k}} (\beta_{\vec{k}} \delta_{\vec{k}} + \beta^+_{\vec{k}} \delta^+_{\vec{k}}) \qquad (8)$$

where $\Gamma_3 = \gamma_-(J_c + Jz)/(u_1 v_1)$, $\Gamma_4 = \gamma_-(J_c + Jz)/(u_2 v_2)$, and $\gamma_- = (\cos k_x - \cos k_y)/2$.

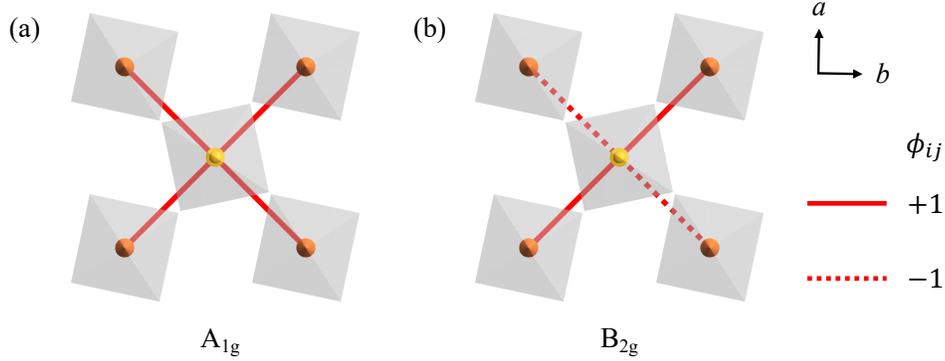

**Fig. S2**    Bond-dependent $\phi_{ij}$ factors of NN bonds in the (a) $A_{1g}$ and (b) $B_{2g}$ scattering Hamiltonian. $\phi_{ij} = +1$ for bonds marked by solid lines, and $-1$ for bonds marked by dashed lines.

The resulting two-magnon scattering cross sections in the $A_{1g}$ and $B_{2g}$ channels are plotted in the inset of the third panel of Fig. 4 of the main text. In the $A_{1g}$ channel, because all lattice symmetries are preserved, the two-magnon cross section captures all DOS features, resulting in two intensity maxima. The lower energy intensity maximum originates from the $\Gamma$ point optical branch, and the higher energy one from the zone edge. In the $B_{2g}$ channel, however, because the four-fold rotational symmetry of the $z$-axis is broken, the contribution from the $\Gamma$ point vanishes. As a result, only one intensity maximum arises from the zone boundary.

### 1.c    Comments on the single layer antiferromagnet

It is worth noting that, if we consider a single layer AFM (*i.e.* $Sr_2IrO_4$), only the intralayer exchange term contributes to the Heisenberg spin Hamiltonian which is identical to and commute with the scattering Hamiltonian in the $A_{1g}$ channel. As a result, this channel does not create any two-magnon excitations unless higher-order terms (*e.g.* beyond NNs) are included in the spin Hamiltonian. In contrast, for the bilayer system that we consider here, $[H, H^S_{A_{1g}}] \neq 0$, and thus two-magnon excitations are allowed in the $A_{1g}$ channel even within the NN approximation.

## 2. Calculations of the beyond nearest-neighbor Heisenberg model of a bilayer antiferromagnet

### 2.a Magnon dispersion

In the simplified model above, only the intralayer NN and interlayer NN exchange coupling $J$ and $J_c$ are considered. In the more realistic model in this section, we further include intralayer second-NN and third-NN exchange interactions $J_2$ and $J_3$ (see Fig. S3) [2], and dipole-like SOC interaction $\Delta$ [3] in the spin Hamiltonian. All of these terms in fact exist in iridates with significant amplitudes when compared to the NN term. In Ref. [3], only the NN SOC interaction has been demonstrated. Here we extend the same type of SOCs to beyond-NN neighbors. Specifically, the SOC term that matters in our case is the second-NN interaction, which is explained in the main text and can be explicitly written as

$$H_{SOC} = \Delta \sum_{i,i',n} (\vec{S}_{i,n} \cdot \vec{\sigma}_2)(\vec{S}_{i',n} \cdot \vec{\sigma}_2) \tag{9}$$

where $\vec{\sigma}_2 = \vec{r}_{i'} - \vec{r}_i$ is a vector pointing from site $i$ to the second-nearest neighbor site $i'$.

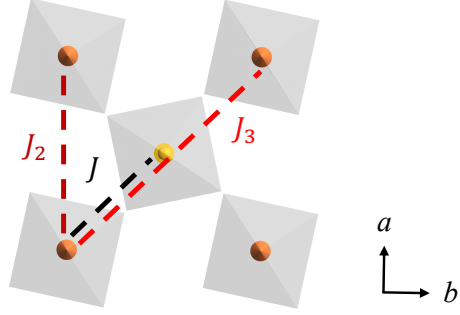

**Fig. S3** Illustration for the in-plane NN, second-NN, and third-NN exchange interactions $J$, $J_2$, and $J_3$.

We apply the Dyson-Maleev transformation in Eq. (2) and the Bogoliubov transformation, leading to

$$\begin{pmatrix}\alpha\\ \gamma\\ \delta\\ \beta\\ \alpha^+\\ \gamma^+\\ \delta^+\\ \beta^+\end{pmatrix} = \frac{1}{4}\begin{pmatrix} \frac{f_1+g_1}{i\sqrt{f_1g_1}} & \frac{-f_1-g_1}{\sqrt{f_1g_1}} & \frac{-f_1-g_1}{\sqrt{f_1g_1}} & \frac{f_1+g_1}{i\sqrt{f_1g_1}} & \frac{-f_1+g_1}{\sqrt{f_1g_1}} & \frac{f_1-g_1}{i\sqrt{f_1g_1}} & \frac{f_1-g_1}{i\sqrt{f_1g_1}} & \frac{-f_1+g_1}{\sqrt{f_1g_1}}\\ \frac{f_2+g_2}{i\sqrt{f_2g_2}} & \frac{-f_2-g_2}{\sqrt{f_2g_2}} & \frac{f_2+g_2}{\sqrt{f_2g_2}} & \frac{-f_2-g_2}{i\sqrt{f_2g_2}} & \frac{f_2-g_2}{\sqrt{f_2g_2}} & \frac{-f_2+g_2}{i\sqrt{f_2g_2}} & \frac{f_2-g_2}{i\sqrt{f_2g_2}} & \frac{-f_2+g_2}{\sqrt{f_2g_2}}\\ \frac{-f_3-g_3}{i\sqrt{f_3g_3}} & \frac{-f_3-g_3}{\sqrt{f_3g_3}} & \frac{f_3+g_3}{\sqrt{f_3g_3}} & \frac{f_3+g_3}{i\sqrt{f_3g_3}} & \frac{f_3-g_3}{\sqrt{f_3g_3}} & \frac{f_3-g_3}{i\sqrt{f_3g_3}} & \frac{-f_3+g_3}{i\sqrt{f_3g_3}} & \frac{-f_3+g_3}{\sqrt{f_3g_3}}\\ \frac{-f_4-g_4}{i\sqrt{f_4g_4}} & \frac{-f_4-g_4}{\sqrt{f_4g_4}} & \frac{-f_4-g_4}{\sqrt{f_4g_4}} & \frac{-f_4-g_4}{i\sqrt{f_4g_4}} & \frac{-f_4+g_4}{\sqrt{f_4g_4}} & \frac{-f_4+g_4}{i\sqrt{f_4g_4}} & \frac{-f_4+g_4}{i\sqrt{f_4g_4}} & \frac{-f_4+g_4}{\sqrt{f_4g_4}}\\ \frac{-f_1+g_1}{\sqrt{f_1g_1}} & \frac{-f_1+g_1}{i\sqrt{f_1g_1}} & \frac{-f_1+g_1}{i\sqrt{f_1g_1}} & \frac{-f_1+g_1}{\sqrt{f_1g_1}} & \frac{-f_1-g_1}{i\sqrt{f_1g_1}} & \frac{-f_1-g_1}{\sqrt{f_1g_1}} & \frac{-f_1-g_1}{\sqrt{f_1g_1}} & \frac{-f_1-g_1}{\sqrt{f_1g_1}}\\ \frac{f_2-g_2}{\sqrt{f_2g_2}} & \frac{f_2-g_2}{i\sqrt{f_2g_2}} & \frac{-f_2+g_2}{i\sqrt{f_2g_2}} & \frac{-f_2+g_2}{\sqrt{f_2g_2}} & \frac{-f_2-g_2}{i\sqrt{f_2g_2}} & \frac{-f_2-g_2}{\sqrt{f_2g_2}} & \frac{f_2+g_2}{\sqrt{f_2g_2}} & \frac{f_2+g_2}{i\sqrt{f_2g_2}}\\ \frac{f_3-g_3}{\sqrt{f_3g_3}} & \frac{-f_3+g_3}{i\sqrt{f_3g_3}} & \frac{f_3-g_3}{i\sqrt{f_3g_3}} & \frac{-f_3+g_3}{\sqrt{f_3g_3}} & \frac{f_3+g_3}{i\sqrt{f_3g_3}} & \frac{-f_3-g_3}{\sqrt{f_3g_3}} & \frac{f_3+g_3}{\sqrt{f_3g_3}} & \frac{-f_3-g_3}{i\sqrt{f_3g_3}}\\ \frac{-f_4+g_4}{\sqrt{f_4g_4}} & \frac{f_4-g_4}{i\sqrt{f_4g_4}} & \frac{f_4-g_4}{i\sqrt{f_4g_4}} & \frac{-f_4+g_4}{\sqrt{f_4g_4}} & \frac{f_4+g_4}{i\sqrt{f_4g_4}} & \frac{-f_4-g_4}{\sqrt{f_4g_4}} & \frac{-f_4-g_4}{\sqrt{f_4g_4}} & \frac{f_4+g_4}{i\sqrt{f_4g_4}}\end{pmatrix}\begin{pmatrix}a\\ b\\ c\\ d\\ a^+\\ b^+\\ c^+\\ d^+\end{pmatrix} \tag{10}$$

where

$$f_1 = \sqrt{J(1+\gamma_{\vec{k}}) - J_2(1-\gamma_{\vec{k}'}) - J_2(1-\gamma_{2\vec{k}}) - \Delta\gamma_{\vec{k}''} + J_c/2},$$

$$g_1 = \sqrt{J(1-\gamma_{\vec{k}}) - J_2(1-\gamma_{\vec{k}'}) - J_2(1-\gamma_{2\vec{k}}) + \Delta\gamma_{\vec{k}''}},$$

$$f_2 = \sqrt{J(1-\gamma_{\vec{k}}) - J_2(1-\gamma_{\vec{k}'}) - J_2(1-\gamma_{2\vec{k}}) + \Delta\gamma_{\vec{k}''} + J_c/2},$$

$$g_2 = \sqrt{J(1+\gamma_{\vec{k}}) - J_2(1-\gamma_{\vec{k}'}) - J_2(1-\gamma_{2\vec{k}}) - \Delta\gamma_{\vec{k}''}},$$

$$f_3 = \sqrt{J(1-\gamma_{\vec{k}}) - J_2(1-\gamma_{\vec{k}'}) - J_2(1-\gamma_{2\vec{k}}) - \Delta\gamma_{\vec{k}''} + J_c/2},$$

$$g_3 = \sqrt{J(1+\gamma_{\vec{k}}) - J_2(1-\gamma_{\vec{k}'}) - J_2(1-\gamma_{2\vec{k}}) + \Delta\gamma_{\vec{k}''}},$$

$$f_4 = \sqrt{J(1+\gamma_{\vec{k}}) - J_2(1-\gamma_{\vec{k}'}) - J_2(1-\gamma_{2\vec{k}}) + \Delta\gamma_{\vec{k}''} + J_c/2},$$

$$g_4 = \sqrt{J(1-\gamma_{\vec{k}}) - J_2(1-\gamma_{\vec{k}'}) - J_2(1-\gamma_{2\vec{k}}) - \Delta\gamma_{\vec{k}''}},$$

and $\gamma_{\vec{k}'} = \cos k_x \cos k_y$, $\gamma_{2\vec{k}} = (\cos 2k_x + \cos 2k_y)/2$, $\gamma_{\vec{k}''} = -\sin k_x \sin k_y$. The eigen-energies are

$$E_1' = 4Sf_1g_1, E_2' = 4Sf_2g_2, E_3' = 4Sf_3g_3, \text{ and } E_4' = 4Sf_4g_4 \tag{11}$$

The intralayer exchange energies of $J = 60$ meV (484 cm$^{-1}$), $J_2 = -20$ meV ($-161$ cm$^{-1}$), and $J_3 = 15$ meV (121 cm$^{-1}$) adopted here are consistent with the previous study of Sr$_2$IrO$_4$ [2]. The interlayer exchange energy $J_c$ is set to 91 cm$^{-1}$, which minimizes the sum of the energy deviations for both M$_1$ and M$_2$. Again $J_c$ is found to be small compared to $J$ (~0.19 of $J$), which supports our choice of the spin wave model. The strength of the SOC $\Delta$ is chosen to be small (16 cm$^{-1}$) so that its value only changes the overall amplitude of the two-magnon feature in the B$_{2g}$ channel. Other qualitative features reported below are insensitive to the value of $\Delta$. We arrive at a spin wave gap of ~ 400 cm$^{-1}$ at the $\Gamma$ point, which is similar to the previous model, and large dispersions along the $X - M$ line, which mimics the case of Sr$_2$IrO$_4$. In this case, the X point is a van Hove singularity point with a divergent DOS. It is only this point that contributes to the DOS at ~ 800 cm$^{-1}$, in contrast to the model above with only NN exchange interaction, where the entire $X - M$ line is responsible for the divergent DOS.

### 2.b Two-magnon cross section

Using the scattering Hamiltonian shown above in Eq. (6), the two-magnon cross section in the A$_{1g}$ channel maintains its shape same, as it does with the NN spin Hamiltonian, with two intensity maxima at ~

800 cm$^{-1}$ and ~ 1700 cm$^{-1}$. However, the van Hove singularity point is invisible to the B$_{2g}$ scattering Hamiltonian with the NN scattering Hamiltonian. Because the second-NN exchange interaction $J_2$ is not negligible (1/3 of $J$) in this more realistic spin Hamiltonian, another B$_{2g}$ scattering Hamiltonian can be constructed with second-NN, which has a checkerboard pattern as shown in Fig. S4. This scattering Hamiltonian takes the form

$$H_{B_{2g}}^{S\prime} = \alpha' \sum_{i,i',n} \phi_{ii'n} (\vec{S}_{i,n} \cdot \vec{S}_{i',n}) \qquad (12)$$

where $\phi_{ii'n} = 1$ for bonds marked by a red solid line and $-1$ for bonds marked by a red dashed line in Fig. S4. This scattering Hamiltonian flips sign under the vertical mirror operation (black dashed line in Fig. S4). Therefore, this scattering Hamiltonian directly couples to the contributions from the van Hove singularity point X. This leads to a cross section which kinks up at ~1700 cm$^{-1}$ as shown in the third panel of Fig. 4 (Please note that the quantum correction factor needs to be considered before comparing to the experimental value of M$_2$).

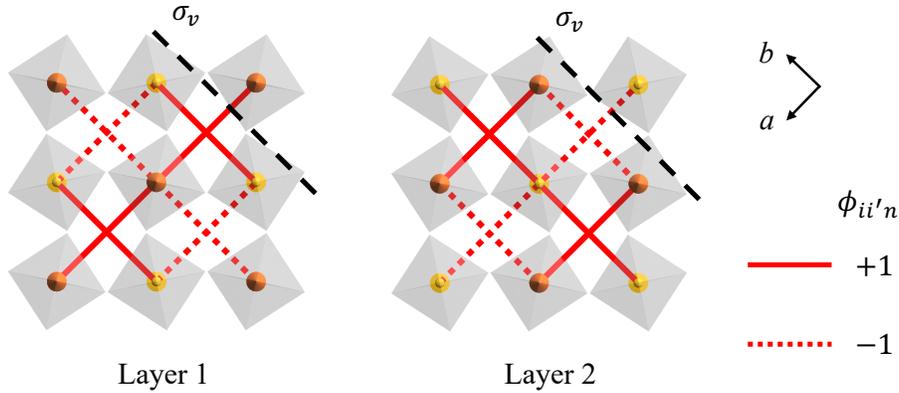

**Fig. S4** Checkerboard pattern of the bond-dependent $\phi_{ii'n}$ factor of second-NN bonds in the new B$_{2g}$ scattering Hamiltonian. $\phi_{ii'n} = +1$ for bonds marked by solid lines, and $-1$ for bonds marked by dashed lines. $\sigma_v$ indicates one of the vertical mirror plane for Sr$_3$Ir$_2$O$_7$.

### 3. Comparison of the two models in Sections 1 and 2

Through the comparison between the two models discussed above, our immediate observation is that the physics of the zone-center two-magnon feature M$_1$ is robust regardless of the model we choose. First, the A$_{1g}$ symmetry ensures that the pair of magnons participating in the two-magnon scattering process must come from the Γ point. Second, this magnon gap energy is not sensitive to the higher-order exchange coupling included in the realistic model in Section 2 which relates spins on two equivalent sites. This is because the Γ point magnons require spins on equivalent sites to have the same orientation, and adding additional neighbors does not affect their energies. To conclude, the simplified model in Section 1 is sufficient in explaining the origin of the zone-center two-magnon feature M$_1$.

We now proceed to the zone-boundary where the other two-magnon feature $M_2$ originates. In the simplified model with only NN exchange coupling in Section 1, the entire zone boundary X-M is dispersionless and contributes to the divergent DOS at the energy of zone-boundary magnons, which leads to the observation of the zone-boundary two-magnon feature $M_2$. When second- and third- NN spin exchanges are included (*i.e.* the realistic model in Section 2), the X point becomes a van Hove singularity point with a divergent DOS while the other momentum points at the zone-boundary do not. For the same reason as in the simplified model, this divergent DOS is responsible for the zone-boundary two-magnon feature. Different from the case of the simplified model, this two-magnon feature arises from two-magnon excitations beyond the NN approximation as mentioned in Section 2.b. This is because every NN term in the $B_{2g}$ scattering Hamiltonian contains cosine pre-factors of $k_x$ or $k_y$. Because $k_x = k_y = \pi/2$ and $\cos k_x = \cos k_y = 0$ at the X point, its contribution to the two-magnon scattering cross section vanishes. Moreover, as the beyond-NN exchange interaction is not negligible in our realistic spin Hamiltonian in Section 2, it is natural to include second-NN in the two-magnon scattering Hamiltonian when the NN contribution vanishes. These terms contain cosines of $k_x + k_y$ and $k_x - k_y$ which survive at the X point. Therefore, the divergent DOS associated with the X point becomes visible in this scattering channel and results in the zone-boundary two-magnon feature $M_2$. In addition, as mentioned in the main text, in the realistic model in Section 2, where the second-NN plays an important role in two-magnon scattering, the dipole-like exchange interaction from SOC becomes a necessary ingredient.

In summary, although the $M_2$ feature arises in both models with similar energies, they in fact have quite different origins. The first model describes better the situation where beyond-NN exchanges are negligible, similar as in the case of cuprates [4], while the second model offers a realistic and interesting manifestation of both the large magnon dispersions along the zone-boundary $X - M$ line as established in the single layer $Sr_2IrO_4$ and the strong SOC in perovskite iridates.

**References**

[1] C. M. Canali and S. M. Girvin, *Theory of Raman scattering in layered cuprate materials*, Physical Review B **45**, 7127 (1992).

[2] J. Kim *et al.*, *Magnetic excitation spectra of Sr2IrO4 probed by resonant inelastic x-ray scattering: establishing links to cuprate superconductors*, Phys Rev Lett **108**, 177003 (2012).

[3] G. Jackeli and G. Khaliullin, *Mott insulators in the strong spin-orbit coupling limit: from Heisenberg to a quantum compass and Kitaev models*, Phys Rev Lett **102**, 017205 (2009).

[4] R. Coldea, S. M. Hayden, G. Aeppli, T. G. Perring, C. D. Frost, T. E. Mason, S. W. Cheong, and Z. Fisk, *Spin waves and electronic interactions in La2CuO4*, Phys Rev Lett **86**, 5377 (2001).